\documentclass[prl,twocolumn,showpacs]{revtex4}
\usepackage{epsfig}
\begin{document}

\author{A. Rebenshtok, E. Barkai}
\affiliation{Department of Physics,
Bar Ilan University, Ramat-Gan 52900 Israel}

\title{Distribution of Time-Averaged Observables for Weak Ergodicity Breaking}


\begin{abstract}

We find a general formula for  
the distribution of time-averaged observables 
for systems modeled according to
the sub-diffusive continuous time random walk.
For Gaussian random walks coupled to a thermal bath we recover 
ergodicity and Boltzmann's statistics, while for the
anomalous subdiffusive case a weakly
non-ergodic statistical mechanical framework is
constructed, which is based on L\'evy's generalized central limit theorem.
As an example we calculate the distribution of $\overline{X}$: the
time average of the position of the particle, for unbiased and uniformly
biased particles, and show that $\overline{X}$ exhibits large
fluctuations compared with the ensemble average $\langle X \rangle$.

\end{abstract}
\maketitle

 A central pillar of statistical mechanics is the ergodic hypothesis;
which yields the equivalence of time and ensemble averages 
in the limit of long measurement time $t$.
The Deborah number $D_{{\rm e}} = t_p/t$ is 
the ratio of the time scale of relaxation of the physical phenomenon under observation $t_p$
and the time of observation \cite{Gupta}.
For a system to exhibit ergodic behavior $D_{{\rm e}}$ must be small.
Recently there is much interest in
weak ergodicity breaking \cite{WEB},
where  the Deborah number  
diverges \cite{Marg,Bel,Euro,Goy,Lomholt}. 
Weakly non-ergodic behavior is found in systems whose dynamics
are characterized by power law distributed   sojourn times in micro-states
of the system,  
in such a way that the averaged waiting
time is infinite (i.e. scale free dynamics). 
Weak ergodicity breaking was investigated
for blinking  quantum dots \cite{Marg}, 
intermittent nonlinear maps generating sub-diffusion deterministically
\cite{Euro}, numerical
simulations of fractional transport in  a washboard potential 
\cite{Goy} and
in vivo gene regulation by DNA-binding proteins \cite{Lomholt}.
On the stochastic level,
all these systems are modeled using the well
known continuous time random walk (CTRW) approach 
or the corresponding fractional Fokker-Planck equation 
\cite{Klages,BouchaudREV,Metzler,Soko}.  
Previously, non-trivial statistics of
occupation times for the CTRW  model were found, and it is well established
that time averages remain 
random variables even in the limit
of long measurement time \cite{Bel}. 
The main open theoretical challenge
is to find 
the distribution of time averages of physical observables. 
Such a general theory, presented in this manuscript,
gives analytical estimates for the statistical deviations
of time averages from ensemble averages. The theory replaces
standard ergodic statistical mechanics, and is applicable for
a wide
class of systems modeled using the CTRW or the related
fractional Fokker-Planck equation.

 We consider the
one-dimensional CTRW on a lattice, with lattice points
$x=1,\cdots,L$. After waiting the particle can jump to one
of its nearest neighbors, with probability $q_x$ it jumps to its
left, and with probability $1-q_x$ to its right.
 The waiting times on lattice cells are independent
identically distributed random variables with a common probability
density function (PDF) $\psi(\tau)$. We consider the widely applicable
case 
\cite{Klages,BouchaudREV,Metzler,Soko,Grig},
where
the PDF of the waiting times behaves like
$ \psi(\tau) \sim  A_\alpha \tau^{ - 1 - \alpha} / | \Gamma( - \alpha)|$
with $0<\alpha<1$, $A_\alpha>0$ when $\tau \to \infty$. In this case
the average waiting time is infinite and the Deborah number diverges. 
Such waiting times  yield anomalous sub-diffusion
and are well investigated \cite{Klages,BouchaudREV,Metzler,Soko,Grig},
in the context of chaotic dynamics \cite{KZS},
geophysics \cite{Berko},
sub-diffusive chemical reactions which are important 
in biological applications \cite{Lindenberg},
and charge transport in amorphous semi-conductors \cite{Scher},
to name a few examples. The vast literature on the CTRW, 
deals mainly with ensemble averages of physical observables,
for example the behavior of the ensemble average
of  the coordinate $\langle X \rangle$ was thoroughly investigated 
in many physical situations. 
Here we investigate the time averages, for example we will find
the distribution of  $\overline{X}$.
 
 Two types of CTRWs are considered. Thermal random walks describe a physical
situation where the particle is coupled to a thermal heat 
bath with a temperature $T$ \cite{BouchaudREV,Metzler}.
In this case the jump probabilities $q_x$
satisfy
usual detailed balance conditions which
relate $q_x$ with an external force field $F(x)$ acting on
the system, 
and temperature $T$ \cite{Bel,BouchaudREV,Metzler}. 
When these conditions are imposed
on the dynamics an ensemble of non-interacting particles attains Boltzmann
equilibrium. A second class of random walks is non-thermal and
this situation may describe a system far from thermal equilibrium.
In this case the ensemble reaches an equilibrium which depends
of-course on the transition probabilities $q_x$ 
(see details below).  
We will treat the non-ergodicity for both cases. 

 We introduce two types of measurements which we identify
with two different types of ensembles. In the first
the time average of a physical observable is made for a
fixed time $t$ and $t \to \infty$. Repeating the experiment
many  times, 
on an ensemble of trajectories,  
the distribution
of the time average is constructed. In the second 
approach the number of jumps $n$ the particle makes is
fixed and $n \to \infty$. So in the first ensemble,
time is fixed and $n$ fluctuates, while the opposite
situation describes the second case. The fixed $n$ fluctuating
time ensemble is very convenient for calculations, and as we discuss
below yields the same results as the fixed time fluctuating 
$n$ approach.  

 We begin the analysis by considering the random
walk where $n$ is the operational time. 
 The probability of occupying lattice site $x$ after $n$ jumps is given
by the discrete time master equation
\begin{equation}
P_x \left(n+1\right) = q_{x+1}P_{x+1} \left( n \right) + \left( 1 -  q_{x-1} \right) P_{x-1} \left( n \right).
\label{eq02}
\end{equation} 
After many jumps $n\to \infty$ an equilibrium $P_{x} ^{{\rm eq}}(n+1)= P_{x} ^{{\rm eq}} (n)$  is obtained, which satisfies
\begin{equation}
P_x ^{{\rm eq}}  = q_{x+1}P_{x+1} ^{{\rm eq}}  + \left( 1 -  q_{x-1} \right) P_{x-1} ^{{\rm eq}}.
\label{eq03}
\end{equation}
Such an equilibrium does not depend on the initial condition
of the system \cite{remark}, 
and is reached provided that the system
is finite, and that $q_x \ne 1$ $q_x \ne 0$ besides on the boundaries.

 We consider the number ensemble where $n$ is fixed. The time
$t_x$ spent by the particle in lattice cell $x$ is called the occupation
time. The total measurement time is $t=\sum_{x=1}^L t_x$.
According to the CTRW model the time $t_x$ is a sum of independent
identically distributed sojourn times with the common power law tailed
PDF $\psi(\tau)$. Let $n_x$
be the number of  sojourn times in cell $x$, which 
is clearly large when $n \to \infty$. For the 
discrete time random walk described by Eq. (\ref{eq02})
we have $n_x/n =P_{x} ^{{\rm eq}}$. 
Hence we may use L\'evy's generalized central limit theorem and 
obtain the PDF of $t_x$
\begin{equation}
f\left(t_x\right) = 
l_{\alpha, A_\alpha P_{x} ^{{\rm eq}} n } \left( t_x \right),
\label{eq04}
\end{equation}
where  
the one sided L\'evy PDF Eq. (\ref{eq04}) is in Laplace $t_x \to u_x$ space
$\hat{f}(u_x) = \exp( - A_{\alpha} P_x ^{{\rm eq}} n u_x ^\alpha )$.
For the ergodic case,  $\alpha=1$ in Eq. (\ref{eq04}),  we have 
$f(t_x) = \delta(t_x - P^{{\rm eq}} _x \langle \tau \rangle n )$,  
where $\langle \tau \rangle = A_1$ is the averaged waiting time,
and since $n \langle \tau \rangle  \to t$
we have $f(t_x) = \delta(t_x - P^{{\rm eq}} _x t )$,  
as expected. 

 The time average of a physical observable $\overline{{\cal O}}$ is 
\begin{equation}
\overline{{\cal O}} = \sum_{x=1, L} \overline{p}_x {\cal O}_x,
\label{eq05}
\end{equation}
where $\overline{p}_x=t_x/t$ is the occupation fraction and ${\cal O}_x$
is the value of the physical observable when the particle is in state
$x$. For example if the observable ${\cal O}$
is the position $X$ of the particle
we have $\overline{X} = \sum_{x=0} ^L x \overline{p}_x$.  For usual ergodic
systems and in the long time limit $\overline{p}_x=P_x ^{{\rm eq}}$
and then the time average is equal to the ensemble average
$\overline{{\cal O} }=
\langle {\cal O} \rangle = \sum_{x=1, L} P_x ^{{\rm eq}} {\cal O}_x$.  
When $\alpha<1$ the dynamics is non-ergodic and $\overline{{\cal O}}$
is a random
variable, even in the long time limit. 

 To obtain the distribution of $\overline{{\cal O}}$ we find now the $L$
dimensional  joint
PDF of the occupation fractions
$P_L(\overline{p}_1,\cdots \overline{p}_x, \cdots \overline{p}_L)$ 
\cite{Barlow}.
First note 
that the $L$ occupation fractions $\overline{p}_x$  
are
constrained
according to the condition
$\sum_{x=1} ^L \overline{p}_x = 1$, hence 
\begin{equation}
 P_L \left( \overline{p}_1, \cdots, \overline{p}_{L}\right)
= \delta\left(1 - \sum_{x=1} ^L \overline{p}_x \right)
\int_0 ^\infty g \left( \overline{p}_1, \cdots \overline{p}_{L-1}, t \right)
{\rm d} t, 
\label{eq06}
\end{equation}
where $g\left(\overline{p}_1, \cdots, \overline{p}_{L-1}, t \right)$
is the $L$ dimensional joint PDF of the random variables in its parenthesis. 
Since the occupation times $t_x$ are all independent we have
$$ g \left( \overline{p}_1, \cdots, \overline{p}_{L-1} , t\right)= $$
\begin{equation}
{ \partial \left( t_1, \cdots,t_{L-1},t\right) \over \partial \left( \overline{p}_1, \cdots,\overline{p}_{L-1}, t \right) } \left[
\Pi_{x=1} ^{L-1} l_{\alpha, A_\alpha P_x ^{{\rm eq}} n } \left(t_x \right) \right] l_{\alpha, A_\alpha P_L ^{{\rm eq}} n } \left( t - \sum_{x=1} ^{L-1} t_x\right).
\label{eq07}
\end{equation}
Calculating the Jacobian we obtain 
$g \left( \overline{p}_1, \cdots, \overline{p}_{L-1} , t\right)$,
then using
Eq. (\ref{eq06}) and the identity
\begin{equation}
l_{\alpha, A_\alpha P_x ^{{\rm eq}} n} ( t_x) ={1 \over (A_\alpha n)^{1/\alpha}}  l_{\alpha,P_x ^{{\rm eq}} }\left( {t_x\over (A_\alpha n)^{1/\alpha}} \right)
\label{eq07a}
\end{equation}
we find 
\begin{equation}
P_L \left( \overline{p}_1 , \cdots, \overline{p}_L\right) = 
\delta\left( 1 - \sum_{x=1}^L \overline{p}_x \right)\int_0 ^\infty {\rm d} y y^{L-1} \Pi_{x=1} ^L l_{\alpha, P_x ^{{\rm eq}} } \left(y \overline{p}_x \right).
\label{eq08}
\end{equation}
This Eq.  is the key for the calculation of the distribution
of the time average $\overline{{\cal O}}$, as we will soon show. 
The multi-dimensional PDF 
of the occupation fractions 
Eq. (\ref{eq08})
is independent of
the number of steps $n$, and the detailed shape of
the waiting time 
besides $\alpha$ of-course
(e.g.  $A_\alpha$ is not important). 
A derivation of Eq. (\ref{eq08}) using the
fixed time ensemble will be presented in a longer publication.   

 To proceed we investigate the 
characteristic function
$ \langle e^{ - u \sum_{x=1} ^L {\cal O}_x t_x} \rangle_t $ 
of the random variable $\sum_{x=1} ^L {\cal O}_x t_x$ in
Laplace $t \to s$ space 
\begin{equation}
 \langle e^{ - u \sum_{x=1} ^L {\cal O}_x t_x} \rangle_s= \int_0 ^\infty e^{ - s t} \langle e^{ - u \sum_{x=1} ^L {\cal O}_x t_x} \rangle_t {\rm d} t. 
\label{eqddollar2}
\end{equation}
Using Eq. (\ref{eq08}) we obtain
\begin{widetext}
$$ \langle e^{ - u \sum_{x=1} ^L {\cal O}_x t_x} \rangle_s = 
\int_0 ^\infty {\rm d} t \int_0 ^\infty {\rm d} y \int_0 ^\infty {\rm d} t_1 \cdots \int_0 ^\infty {\rm d} t_L t \delta\left( t - \sum_{x=1} ^L t_x \right) y^{L-1}  e^{ - s t - u \sum_{x=1} ^L {\cal O}_x t_x } \Pi_{x=1} ^L l_{\alpha, P_x ^{{\rm eq}} } \left( y t_x  \right) = $$
$$ - { {\rm d} \over {\rm d} s} \int_0 ^\infty {\rm d} y y^{L-1} \int_0 ^\infty {\rm d} t_1 \cdots \int_0 ^\infty {\rm d} t_L e^{ - s \sum_{x=1} ^L t_x - u \sum_{x=1} ^L {\cal O}_x t_x} \Pi_{x=1}  ^L l_{\alpha,P_x ^{{\rm eq}}}\left(y t_x \right) =  $$
\end{widetext}
\begin{equation}
- {{\rm d} \over {\rm d} s } \int_0 ^\infty {\rm d} y y^{L-1} \Pi_{x=1}^L 
\left\{  {\exp\left[ - P_x ^{{\rm eq}} \left( { s +  {\cal O}_x u \over y} \right)^\alpha \right]\over y} \right\}.
\label{eq09}
\end{equation}
Solving the last integral we find the characteristic function
\begin{equation}
\langle e^{ - u \sum_{x=1} ^L {\cal O}_x t_x} \rangle_s =   { \sum_{x=1} ^L P_{x} ^{{\rm eq}} \left( s +  {\cal O}_x u\right)^{\alpha - 1} \over \sum_{x=1} ^L P_{x} ^{{\rm eq}} \left( s +  {\cal O}_x u \right)^\alpha}.  
\label{eq10}
\end{equation}
Using inversion technique found in \cite{Godreche}, we transform 
Eq. (\ref{eq10}),  and find the PDF
of the time average $\overline{{\cal O}}$ 
\begin{equation} 
f_{\alpha} \left( \overline{{\cal O}} \right) = 
- { 1 \over \pi} \lim_{\epsilon\to 0} \mbox{Im} 
{ \sum_{x=1} ^L P^{{\rm eq}} _x \left( \overline{{\cal O}} - {\cal O}_x + i \epsilon\right)^{\alpha -1} \over
 \sum_{x=1} ^L P^{{\rm eq}} _x \left( \overline{{\cal O}} - {\cal O}_x + i \epsilon\right)^\alpha}.
\label{eq11}
\end{equation}
This is our main result so far, it is a very general formula
for the distribution of time-averaged observables 
and is valid for a CTRW
on a lattice. 
In the limit $\alpha \to 1$ 
\begin{equation}
f_{\alpha=1} \left( \overline{{\cal O}} \right) = \delta \left( \overline{{\cal O}} - \langle {\cal O} \rangle \right)
\label{eq12}
\end{equation}
which is the expected ergodic behavior. 
The opposite limit of $\alpha \to 0$ gives
\begin{equation}
\lim_{\alpha \to 0} f_{\alpha} \left( \overline{{\cal O}} \right) =
\sum_{x=1} ^L P^{{\rm eq}} _x \delta\left( \overline{{ \cal O}} - {\cal O}_x \right). 
\label{eq13}
\end{equation}
This makes perfect physical sense, since when $\alpha \to 0$ the particle
is localized for the whole duration of measurement in a single cell 
\cite{remark1}. 
Note that our results can be easily generalized to dimensions higher than 
one. 

In many applications the continuum behavior of the CTRW is important.
Dynamically this limit corresponds to the behavior described
by the fractional time Fokker--Planck equation \cite{Metzler,Soko}.
Taking the
continuum limit of  Eq. (\ref{eq11}) we find
\begin{equation}
f_\alpha\left( \overline{{\cal O}} \right)  = - {1 \over \pi} \lim_{\epsilon \to 0 } \mbox{Im} { \int_0 ^L {\rm d} x P^{{\rm eq}} \left(x\right) \left[ \overline{{\cal O}} - {\cal O}(x) + i \epsilon \right]^{\alpha -1} \over
\int_0 ^L {\rm d} x P^{{\rm eq}} \left( x \right) \left[
\overline{{\cal O}} - {\cal O}(x)  + i \epsilon \right]^\alpha }.
\label{eq14}
\end{equation}
Here $P^{{\rm eq}} \left( x \right){\rm d} x$ is the equilibrium
probability 
(in ensemble sense) 
of finding the particle in $(x,x+ {\rm d} x)$ and $0<x<L$.
When
the random walk is coupled to a thermal heat bath with temperature
$T$, in the presence of an external force field $F(x)$,
the equilibrium of the ensemble is described by Boltzmann's statistics 
\cite{BouchaudREV,Metzler}
\begin{equation}
P^{{\rm eq}} (x) = {\exp\left[ - {V\left(x\right) \over k_b T} \right] \over Z},
\label{eq15}
\end{equation}
where $Z$ is the partition function and $F(x) = - {\rm d} V(x)/ {\rm d} x$.
As mentioned such an equilibrium is found for the CTRW model 
when detailed balance conditions
are imposed on $q_x$. 
Solving Eq. (\ref{eq14}) we have 
$$ f_{\alpha} \left( \overline{{\cal O}} \right) =  $$
\begin{equation}
{ \sin \pi \alpha \over \pi} 
{I^{<} _{\alpha - 1} \left( \overline{{\cal O}} \right) I^{>} _{\alpha}\left( \overline{{\cal O}} \right) + I^{>} _{\alpha -1} \left( \overline{{\cal O}} \right)  I^{<}_{\alpha} \left( \overline{{\cal O}} \right)\over
\left[ I^{>} _{\alpha } \left( \overline{{\cal O}} \right)\right]^2 +
\left[ I^{<} _{\alpha } \left( \overline{{\cal O}} \right)\right]^2 +
2 \cos \pi \alpha I^{>} _{\alpha } \left( \overline{{\cal O}} \right) I^{<} _{\alpha } \left( \overline{{\cal O}} \right) },
\label{eq16}
\end{equation}
where
\begin{equation}
I^{<} _{\alpha } \left( \overline{{\cal O}} \right)=
\int_{\overline{{\cal O}} < {\cal O} (x)} {\rm d} x P^{{\rm eq}} (x) | \overline{{\cal O}} - {\cal O} (x)|^{\alpha} 
\label{eq17}
\end{equation}
and similarly for $I^{>}_{\alpha}\left( \overline{{\cal O}} \right)$,
$I^{<}_{\alpha-1}\left( \overline{{\cal O}} \right)$
and  $I^{>}_{\alpha-1}\left( \overline{{\cal O}} \right)$.
The integration domain in Eq. (\ref{eq17}) is for $x$ satisfying
the condition $\overline{{\cal O}} < {\cal O} (x)$. 

\begin{figure}
\begin{center}
\epsfxsize=80mm
\epsfbox{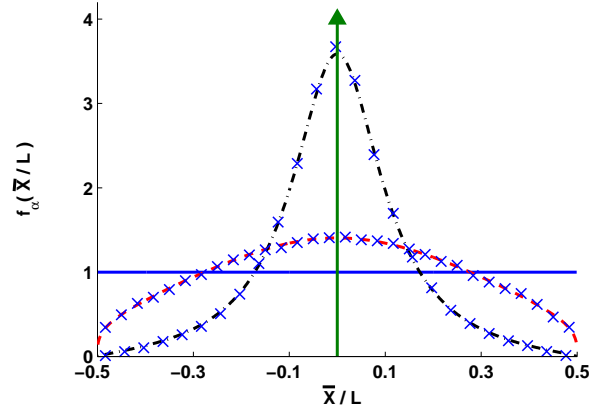}
\end{center}
\caption{ 
The PDF of $\overline{X}/L$ for unbiased CTRW, simulations (crosses)
 versus theory (curves) Eq. (\ref{eq18}). 
 When $\alpha=1$ we find
ergodic behavior and $\overline{X} = \langle X \rangle =0$ 
(i.e. the arrow symbolizing a delta function). 
For $\alpha=0.7$ (the dotted dashed curve) and $\alpha =0.3$  (the dashed
curve) large fluctuations of time averages are observed. When
$\alpha \to 0$ the PDF of $\overline{X}$ is uniform reflecting 
localization of the particle (solid line).    
}
\label{fig1}
\end{figure}

 As an example consider a particle in a domain $-L/2<x<L/2$ undergoing
an unbiased random walk. This is a free particle in the sense that
no external field is acting on it. The time average of the particle's
position $\overline{X}$ is considered, and obviously for this
case $P^{{\rm eq}}(x) = 1/L$ for $-L/2<x<L/2$. Using Eq. 
(\ref{eq16}) we find the PDF of the time-averaged position
$$ f_{\alpha} \left( \overline{X} \right) = $$
\begin{equation}
{ 1 \over L}  
{ N_\alpha  \left( { 1 \over 4} - { \overline{X}^2 \over L^2} \right)^\alpha  \over
\left| { 1 \over 2} - { \overline{X} \over L} \right|^{ 2 ( 1 + \alpha) } +
\left| { 1 \over 2} + { \overline{X} \over L} \right|^{ 2 ( 1 + \alpha) } +
2 \left| { 1 \over 4} -  \left( { \overline{X} \over L} \right)^2 \right|^{  1 + \alpha } \cos \pi \alpha }.
\label{eq18}
\end{equation}
where $N_\alpha=( 1 + \alpha ) \sin \pi \alpha / (\pi \alpha )$.
When $\alpha \to 1$ we have the  ergodic behavior
$\overline{X} = \langle X \rangle=0$ while
$f_{\alpha \to 0} \left( \overline{X} \right) = 1/L $ for $|\overline{X}|<L/2$
which is the uniform distribution, reflecting the mentioned
localization of the
particle in space when $\alpha \to 0$. 
In Fig. 
\ref{fig1} comparison between our analytical results and
numerical simulations \cite{remark2} of the
CTRW process with a fixed measurement time $t$,  
show excellent agreement without fitting. 

\begin{figure}
\begin{center}
\epsfxsize=80mm
\epsfbox{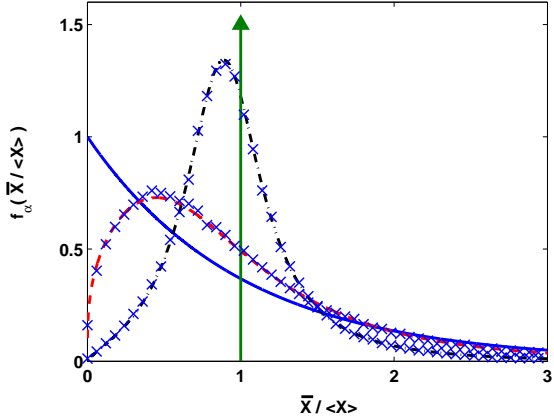}
\end{center}
\caption{ 
Same as Fig. \ref{fig1} for biased CTRW:  now we show the PDF
of $\overline{X}/\langle X \rangle$.
The theoretical curves based on
Eq. (\ref{eq19})  perfectly agree
with simulations (crosses) without fitting.
A transition between ergodic behavior for $\alpha=1$ (the delta function)
to localization behavior (solid curve $\alpha \to 0$)
where the PDF of $\overline{X}$ decays exponentially
is found.
}
\label{fig2}
\end{figure}

 As a second example consider a biased particle
in the domain $0<x<\infty$ 
and  in a constant force
field $F>0$.  
Assuming the particle is in contact with a heat bath, with temperature
$T$, Boltzmann's equilibrium is reached for an ensemble of
particles, $P^{{\rm eq}} (x) = \exp( - F x / k_b T)/Z$.
The PDF of  $\overline{X}$
is found using Eq.  
(\ref{eq16}) 
$$ f_{\alpha} \left( \overline{X} \right) = 
{ \sin \pi \alpha \over \pi} { F \over k_b T} \times $$
\begin{equation}
{ \Gamma\left(\alpha \right) e^{\tilde{x}} \tilde{x}^\alpha
 \over
\left( \int_0 ^{\tilde{x}} {\rm d} y e^y |y|^\alpha\right)^2 + \Gamma^2 (1 + \alpha) + 2 \Gamma(1 + \alpha)  \int_0 ^{\tilde{x}} {\rm d} y e^y |y|^\alpha \cos \pi \alpha }, 
\label{eq19}
\end{equation}
where $\tilde{x} = F \overline{X} / k_b T$.
When $\alpha\to 1$ we find ergodicity $f_1(\overline{X}) = \delta(\overline{X} - \langle X \rangle)$ with $\langle X \rangle= k_b T / F$ while 
in the opposite limit $\alpha \to 0$,
an exponential decay of the PDF of $\overline{X}$ is
found: 
$\lim_{\alpha\to 0} f_\alpha(\overline{X}) = \exp( - F \overline{X} / k_b T )/Z$
reflecting localization with a profile determined by the equilibrium
density of many
non-interacting particles. 
These behaviors are demonstrated in Fig. \ref{fig2}. 

 We now discuss briefly the meaning of weak ergodicity breaking.
In many situations
in Physics a system is non-ergodic since its phase space is
decomposed into regions of phase space where the system
once starting in one region cannot explore the others. 
In this case time averages depend strongly on the initial
condition of the system and there is no full exploration
of phase space. In contrast for weak ergodicity breaking, the particle
will visit each lattice cell many times, no matter what is
its initial condition. Hence exploration of phase space is
possible, and for this reason
we were able to construct in this manuscript
a general theory of
non-ergodic statistical mechanics which is not sensitive to 
the initial conditions of the system.
 This has several implications, for
example the joint PDF of occupation fractions Eq.
(\ref{eq08})
and the PDF of time averages Eqs. (\ref{eq14},\ref{eq16})
are related to the population density $P^{{\rm eq}}(x)$.
 Therefore we may
find a general relation between fluctuations of time averages
and fluctuations of ensemble averages: 
using the small $u$ expansion of
Eq.  
(\ref{eq10})
$$ \langle \overline{{\cal O}}^2 \rangle - \langle \overline{{\cal O }} \rangle^2= $$
\begin{equation}
\left( 1 - \alpha\right)\left[ \int_{0} ^{L} {\cal O}\left( x \right)^2 P^{{\rm eq}} 
\left(x\right) {\rm d} x  - 
\left( \int_{0} ^{L} {\cal O} \left( x \right)  P^{{\rm eq}}  \left(x\right) {\rm d} x \right)^2 \right], 
\label{eq20}
\end{equation}
while the average of $\overline{{\cal O}}$ is  
$\langle \overline{{\cal O}} \rangle = \langle {\cal O} \rangle= 
\int_0 ^L {\cal O} P^{{\rm eq}}(x) {\rm d} x$. 
For the example of a particle in a uniform force
field $F$, when the physical observable is the position,
we have $\langle \overline{X} \rangle = k_b T/ F$ and
$\langle \overline{X}^2 \rangle - \langle \overline{X} \rangle^2 = ( 1 - \alpha) (k_b T/F)^2 $.   

 To summarize we have obtained  very general
distributions of time averages
of physical observables of weakly non-ergodic systems Eqs. (\ref{eq11},\ref{eq16}).
Unlike usual ergodic statistical mechanics
where the time averages are equal to the ensemble averages,
we find large fluctuations of time averages. 
Due to the large number of applications of the CTRW  model,
and the recent interest in weak ergodicity breaking in dynamics of
single particles, our theory is likely to find its applications
in many systems. Due to the deep relations between the stochastic
CTRW model and
other models of anomalous diffusion, e.g. the quenched trap model, 
and deterministic dynamics,      
our non-ergodic theory might find further profound justification.   

{\bf Acknowledgment:} This work was supported by the 
Israel Science Foundation.


\begin{thebibliography}{99}
\bibitem{remark} For reflecting boundary conditions,
it is easy to show that $P^{{\rm eq}}_x = 
N^{-1} [1/(1-q_x)]\Pi_{k=2} ^x (1-q_k)/q_k$  with the normalization
$N = \sum_{x=1} ^L [1 / ( 1 - q_x) ]  \Pi_{k=2}^x (1-q_k)/q_k$.
\bibitem{remark1} 
We take the limit $t \to \infty$ before $\alpha \to 0$, otherwize
the particles cannot reach equilibrium.
\bibitem{remark2} In simulations we used: $\psi(\tau)= \alpha \tau^{-1-\alpha}$ when $\tau>1$ otherwise $\psi(\tau)=0$, $L=30$, $120000$
realizations, and the
maximal measurement time (for $\alpha=0.3$) was $t=10^{16}$.
For the biased CTRW shown in
Fig. \ref{fig2} we have $\langle X \rangle = 16.7$ (lattice spacing
is unity).
\end{thebibliography}
\end{document}